\documentclass{scrartcl}

\usepackage[english]{babel}
\usepackage{amsmath}
\usepackage{bbm}
\usepackage{slashed}
\usepackage{parskip}

\newcommand{\eqnl}[2]{\begin{eqnarray}#1\label{#2}\end{eqnarray}}
\newcommand{\eqnn}[1]{\begin{eqnarray*}#1\end{eqnarray*}}

\newcommand{\eqngr}[2]{\begin{eqnarray*}#1\\#2\end{eqnarray*}}

\newcommand{\eqngrl}[3]{\begin{eqnarray}#1\nonumber\\#2\label{#3}\end{eqnarray}}

\newcommand{\eqngrrl}[4]{\begin{eqnarray}#1\nonumber\\#2\label{#4}\\#3\nonumber\end{eqnarray}}

\newcommand{\N}{\mathbbm{N}}
\newcommand{\R}{\mathbbm{R}}
\newcommand{\C}{\mathbbm{C}}
\newcommand{\id}{\mathbbm{1}}

\newcommand{\mN}{\mathcal{N}}
\newcommand{\mM}{\mathcal{M}}
\newcommand{\mH}{\mathcal{H}}
\newcommand{\mF}{\mathcal{F}}

\newcommand{\ud}{{\textrm{d}}}
\newcommand{\ui}{{\textrm{i}}}
\newcommand{\ue}{{\textrm{e}}}
\newcommand{\uB}{{\textrm{B}}}
\newcommand{\uF}{{\textrm{F}}}
\newcommand{\ind}{{\textrm{ind}\;}}
\newcommand{\im}{{\textrm{im}\;}}

\newcommand{\vecbf}[1]{\boldsymbol{#1}}

\newcommand{\pa}{\partial}

\newcommand{\va}{|0\rangle}
\newcommand{\fdi}{\slashed{\partial}}
\newcommand{\di}{\slashed{D}}

\def\mtxt#1{\quad\hbox{{#1}}\quad}
\def\ese{\!&\!=\!&\!}
\def\ms{\!-\!}
\def\es{\!=\!}
\def\ps{\!+\!}

\def\psiA{\psi^A}
\def\psiB{\psi^B}
\def\psidA{\psi^{A\dagger}}
\def\psidB{\psi^{B\dagger}}

\def\refs#1{(\ref{#1})}

\def\In#1#2{I^{#1}_{\hskip2.2mm #2}}
\def\Ini#1#2{I^{#1}_{i\hskip1.5mm #2}}
\def\Jn#1#2{J^{#1}_{\hskip2mm #2}}

\def\PAu{P^A_{\hskip 1.6mm B}}
\def\IMu{I^M_{\hskip 1.6mm N}}

\def\fNu{f^N_{\hskip 2mm \mu}}

\def\fMl{f^\mu_{\hskip 1.5mm M}}
\def\fbMl{f^{\bar\mu}_{\hskip 1.5mm M}}
\def\fNl{f^\mu_{\hskip 1.5mm N}}
\def\fbNl{f^{\bar \mu}_{\hskip 1.5mm N}}
\def\fMu{f^M_{\hskip 2.5mm \mu}}
\def\fbMu{f^M_{\hskip 2.5mm \bar\mu}}

\def\om{\omega}

\def\al{\alpha}

\def\EAu{E_{M}^{A}}
\def\EBu{E_{N}^{B}}
\def\EAl{E_{A}^{M}}
\def\EBl{E_{B}^{N}}

\newcommand{\q}{{\textstyle\frac{1}{4}}}
\newcommand{\h}{{\textstyle\frac{1}{2}}}

\newcommand{\ih}{{\textstyle\frac{\ui}{2}}}
\newcommand{\iq}{{\textstyle\frac{\ui}{4}}}

\begin{document}

\title {Extended Supersymmetries and the Dirac Operator}
\author{A. Kirchberg, J.D. L\"ange and A.
Wipf\footnote{\tt{A.Kirchberg@tpi.uni-jena.de,
J.D.Laenge@tpi.uni-jena.de, A.Wipf@tpi.uni-jena.de}} \\[10pt]
Theoretisch--Physikalisches Institut, \\
Friedrich--Schiller--Universit\"at Jena,\\
07743 Jena, Germany}
\date{}

\maketitle

\vspace{-250pt}
\hfill{FSU-TPI 01/04}
\vspace{+250pt}

\paragraph{Abstract:} We consider supersymmetric quantum mechanical
systems in arbitrary dimensions on curved spaces with
nontrivial gauge fields. The square of the Dirac
operator serves as Hamiltonian. We derive a relation between
the number of supercharges that exist and restrictions on the geometry
of the underlying spaces as well as the admissible gauge field
configurations. From the superalgebra with two or more real
supercharges we infer the existence of
integrability conditions and obtain a corresponding superpotential.
This potential can be used to deform the supercharges
and to determine zero modes of the Dirac operator.
The general results are applied to the K\"ahler spaces $\mathbbm{CP}^n$.  



\section{Introduction}
\label{section1}

Supersymmetry is a crucial ingredient
in many attempts to unify the
interactions contained in the
standard model of particle physics.
It softens the ultraviolet divergences
and offers the hope of resolving
the hierarchy problem. It arises
naturally in low-energy limits of string
theory. Supersymmetric models
are easier to solve than their
non-supersymmetric counterparts, since they are
more strongly constrained by the higher degree of symmetries.

In recent years we have seen a renewed interest
in nonperturbative aspects of strongly interacting
supersymmetric theories. This is mainly due to
the \emph{Seiberg-Witten solution} for the low-energy
effective action of $\mN\es 2$ super-Yang-Mills theory
\cite{seibergwitten} and the \emph{Maldacena conjecture}
stating that $\mN\es 4$ super-conformal
$SU(N_c)$-gauge theories arising on parallel
D3-branes are in the limit of large 't Hooft
coupling and large $N_c$ dual to supergravity
theories on an $AdS_5$-background \cite{maldacena}.
Despite of these striking results
there is still a long way to go towards a better
understanding of nonperturbative
effects in supersymmetric theories with less
supersymmetries and finite $N_c$.
In particular, since low-energy physics is manifestly
not supersymmetric, it is necessary that this
symmetry is broken at some energy scale. As
issues of supersymmetry breaking are difficult to address in
perturbation theory, one is motivated to
study supersymmetric models on a \emph{spacetime lattice}.
Unfortunately, supersymmetry is explicitly broken
by most discretization procedures,
and it is a nontrivial problem to recover supersymmetry
in the continuum limit. However, there
are discretizations with
nonlocal interaction terms for which
supersymmetry is manifestly realized \cite{nonlocalsusy}.
Alternatively, for some models one can
discretize space -- but not time -- such that a
subalgebra of the supersymmetry algebra which
determines spectral properties of
the super-Hamiltonian remains intact \cite{susylattice}.

Every supersymmetric field theory on a spatial lattice
may be reinterpreted as a higher-dimensional
\emph{supersymmetric quantum mechanical system}.
The first studies of
such systems go back to Nicolai \cite{Nicolai:1976xp}
and have been extended by Witten in his
work on supersymmetry breaking
\cite{Witten:1981nf,Witten:1982im,Witten:1982df}. 
Soon after that, de Crombrugghe and Rittenberg \cite{Rittenberg:1983}
presented a very general analysis of supersymmetric
Hamiltonians. 
Over the years, it has been demonstrated
that supersymmetry is a useful technique to
construct exact solutions in quantum
mechanics \cite{qmproblems}. For example,
all ordinary Schr\"odinger equations with
shape invariant potentials can be solved
algebraically with the methods of supersymmetry.
On the other hand, apparently different
quantum systems may be related by supersymmetry,
and this relation may shed new light on the physics
of the two systems. For example, the
hydrogen atom (its Hamiltonian, angular momentum
and Runge-Lenz vector) can be supersymmetrized.
The corresponding theory contains
both the proton-electron and the proton-positron
system as subsectors \cite{klpw}.

The present work contains the first part of
our attempt to better understand supersymmetric
field theories on spatial lattices. Here, we will
analyze properties of quantum mechanical systems. In a forthcoming
publication, our results will be related to Wess-Zumino models
on such lattices. This paper is
organized as follows: In Section 2 we recall
supersymmetric quantum mechanics with $\mN$ supersymmetries.
The main emphasis is on the algebraic
structure of such systems. In the following section
we give explicit realizations of systems
with one, two or more supersymmetries. They
are based on the Dirac operator in external
gauge and gravitational fields. We shall
see that for certain background fields
there are $\mN$ inequivalent ways to take
the square root of $-\di^2$. At the same time
$-\di^2$ commutes with several  particle-number
operators which correspond to complex structures.
The superalgebra implies
consistency conditions for these structures and
the gauge field strength. For example, the
Dirac operator in four dimensions admits an
extended $\mN\es 4$ supersymmetry if spacetime
is hyper-K\"ahler and the gauge field is (anti-)selfdual.
In Section 4 we show that, for background fields
admitting an extended supersymmetry, the
geometry and gauge potential are encoded
in a superpotential. The
superpotential may be used to deform the generally-
and gauge-covariant derivative into the
ordinary derivative.
In Section 5 we apply our general results to
study the Dirac operator on the complex
projective spaces $\C P^n$ with an Abelian
background gauge field. We derive explicit
expressions for the superpotential
and fermionic zero modes on these K\"ahler spaces.

\section{Extended Supersymmetric Quantum Mechanics}
\label{section2}

Supersymmetric quantum mechanics describes systems
with nonnegative Hamiltonians that can be written as
\eqnl{
\delta_{ij}H=\h \left\{Q_i,Q_j \right\},\qquad i,j=1,\ldots,\mN,}{sqm1}
with \emph{Hermitian} supercharges $Q_i$ anticommuting with an
involutary operator $\Gamma$,
\eqnl{
\{ Q_i , \Gamma \}= 0, \qquad \Gamma^\dagger=\Gamma,\qquad
\Gamma^2 =\id . }{sqm2}
There are various
definitions of supersymmetric quantum mechanics in the literature,
for a recent discussion, in particular concerning the role of the
grading operator $\Gamma$, we refer to \cite{Combescure:2004ey}.
One may also relax the
condition for the left-hand side of (\ref{sqm1}), see for example
\cite{Hull:1999ng}, but in this paper we will not consider such
systems.

The $+1$ and $-1$ eigenspaces of $\Gamma$ are called
\emph{bosonic} and \emph{fermionic sectors} respectively,
\eqnl{
\mH =\mH_{\uB} \oplus \mH_{\uF}, \quad \mH_B=\mathcal{P}_+\mH, \quad
\mH_F=\mathcal{P}_-\mH, \quad \mathcal{P}_\pm=\h(\id\pm\Gamma).}{sqm4}
The supercharges $Q_i$
map $\mH_{\uB}$ into $\mH_{\uF}$ and vice versa.
The super-algebra \refs{sqm1} implies that they
commute with the super-Hamiltonian,
\eqnl{
[ Q_i,H ] = 0,}{sqm5}
and generate supersymmetries of the system. The simplest models
exhibiting this structure are $2\times2$-matrix
Schr\"odinger operators in one dimension
\cite{Nicolai:1976xp,Witten:1981nf,Witten:1982im}.
In this paper we shall investigate
explicit representations of the superalgebra
\refs{sqm1} with one, two, four and more supercharges.

\paragraph{One supercharge:}
In this case every eigenstate
of $H=Q_1^2\geq 0$ with positive energy is paired
by the action of $Q_1$. For example, if $|B\rangle$ is a bosonic
eigenstate with positive energy, then
$|F\rangle\sim Q_1 |B\rangle$ is
a fermionic eigenstate with the same
energy. However, a normalizable eigenstate
with zero energy is annihilated by the supercharge,
$Q_1\va =0$, and hence has no superpartner.
In a basis where $\Gamma=\sigma_3\otimes\id$,
the Hermitian charge $Q_1$ has the form
\eqnl{
Q_1 =\mathcal{P}_-Q_1\mathcal{P}_+ + \mathcal{P}_+Q_1\mathcal{P}_-\equiv
\begin{pmatrix} 0 & A^\dagger \\ A & 0 \end{pmatrix}.}{sqm7}
The index of $Q_1$ counts the difference of bosonic
and fermionic zero modes,
\eqnl{
\ind Q_1 = \dim \ker A - \dim \ker A^\dagger
= n^0_\uB - n^0_\uF. }{sqm8}
Supersymmetry is spontaneously broken if and only if
there exists no state which is left invariant by the supercharges,
or equivalently if $0$ is not in the discrete spectrum of $H$.

\paragraph{Two supercharges:}
In this case there exist two anticommuting
and Hermitian roots of the super-Hamiltonian
\eqnl{
H=Q_1^2=Q_2^2,\quad \{Q_1,Q_2\}=0,\quad
Q_i^\dagger=Q_i.}{sqm9}
Later we shall use the
nilpotent \emph{complex supercharge}
\eqnl{
Q = \h (Q_1+\ui Q_2),}{sqm10}
and its adjoint $Q^\dagger$, in
terms of which the supersymmetry algebra takes
the form
\eqnl{
H = \{ Q , Q^\dagger \},\quad Q^2=Q^{\dagger\,2}=0
\mtxt{and}\quad[Q,H]=0.}{sqm12}
The number of normalizable zero modes of $H$ is
given by \cite{Witten:1981nf}
\eqnl{
n^0 = n^0_\uB + n^0_\uF = \dim (\ker Q / \im Q) = \dim (\ker
Q^\dagger / \im Q^\dagger).}{sqm13}
\paragraph{Four supercharges:}
Now there are four distinct roots
of the super-Hamiltonian,
\eqnl{
H=Q_1^2=Q_2^2=Q_3^2=Q_4^2.}{sqm14}
The only nontrivial anticommutators of
the complex nilpotent supercharges
\eqnl{
Q=\h(Q_1+\ui Q_2)\mtxt{and}
\tilde Q=\h (Q_3+ \ui Q_4)}{sqm15}
and their adjoints are
\eqnl{
\{Q,Q^\dagger\}=\{\tilde Q,\tilde Q^\dagger\}=H.}{sqm16}

\section{Supersymmetries and the Euclidean Dirac Operator}

There exists a fundamental supersymmetric
Hamiltonian in nature,
the square of the Euclidean Dirac operator. In other
words, one identifies the Dirac operator
as a supercharge associated to this Hamiltonian.
There are non-linear sigma-models which give rise to
exactly these supercharges, in particular the
$(1+0)$-dimensional models studied in \cite{Hull:1999ng, Coles:1990hr}.
In contrast to those models, we allow for the presence
of gauge fields but do not include torsion.
The identification of Dirac operators and supercharges
has also been employed by Alvarez-Gaum\'e in \cite{Alvarez-Gaume:1983at},
where he uses supersymmetry to derive the Atiyah-Singer
index theorem.

The \emph{chiral supersymmetry} with one charge exists
in all even dimensions and for arbitrary
gauge and gravitational background fields.
It can be extended if $\di^2$ commutes with certain
\emph{particle-number operators} to be defined below.
For example, in $D\es 2n$ Euclidean dimensions and for
background fields
with \emph{holonomy group} U$(n)$ the operator
$\di^2$ commutes with one particle-number
operator and admits two supersymmetries. In $D\es 4n$
dimensions and for background fields with
\emph{holonomy group} Sp$(n)$, there are
three conserved number operators and
four supersymmetries. 

We consider a smooth Riemannian manifold $\mM$ of
dimension $D$ which allows for a spin structure.
We describe the gravitational fields in terms
of vielbeins $\EAu$ rather than
a metric $G_{MN}$, which is related to the vielbein by
\eqnl{
G_{MN} = \EAu \EBu \delta_{AB},\qquad
\delta^{AB}=G^{MN}\EAu\EBu.}{dirop1}
The Lorentz indices $A,B\in\{1,\dots,D\}$ are converted into
coordinate indices $M,N\in\{1,\dots,D\}$ (or vice versa) with the
help of the vielbein $\EAu$ or its inverse, which is given by 
$\EAl=G^{MN}\EBu \delta_{BA}$.
The Clifford algebra is generated by the Hermitian
matrices $\Gamma^A$, satisfying
\eqnl{
\{ \Gamma^A , \Gamma^B \}=2 \delta^{AB}\mtxt{or}
\{ \Gamma^M , \Gamma^N \}=2 G^{MN},}{dirop2}
where the $\Gamma^M=\Gamma^A\EAl$ are the matrices
with respect to the holonomic basis $\pa_M$.

\subsection{Chiral Supersymmetry}

In even dimensions we always have
\emph{chiral supersymmetry}
generated by the Hermitian Euclidean Dirac
operator, viewed as supercharge 
\eqnl{
Q_1=\ui \di
=\ui\Gamma^M\nabla_M=\ui \Gamma^A\nabla_A, \quad \nabla_A=\EAl\nabla_M.}
{dirop4}
The generally- and gauge-covariant derivative
acting on spinors,
\eqngrl{
\nabla_M\ese \pa_M + \Omega_M+A_M}
{\ese \pa_M+\q \Omega_{MAB}\Gamma^{AB}+A_M^a T_a,}{dirop5}
contains the connection $\Omega$ and gauge potential $A$ together
with the anti-Hermitian generators
$\Gamma^{AB}\es \h \left[\Gamma^A,\Gamma^B\right]$ and $T^a$
of spin rotations and gauge transformations.
The gamma-matrices are covariantly
constant in the following sense,
\eqnl{
\nabla_M\Gamma^N=\pa_M\Gamma^N+\Gamma^{N}_{MP}\Gamma^P
+[\Omega_M,\Gamma^N]=0.}{dirop3}
For the involutary operator $\Gamma$ in \refs{sqm2} we take
in $D\es 2n$ dimensions
\eqnl{
\Gamma =\al \Gamma^1 \ldots \Gamma^D,}{dirop7}
where the phase $\al$ is chosen such
that $\Gamma$ is Hermitian and squares to $\id$,
$\al^2=(-)^n$. The `bosonic' and `fermionic' subspaces
consist of spinor fields with positive and negative
chiralities, respectively, and the number of bosonic
minus the number of fermionic zero
modes equals the index of the Dirac operator,
\eqnl{
n^0_\uB - n^0_\uF =\ind \ui \di \;. }{dirop8}

Since the commutator of two covariant derivatives
yields the gauge field strength and curvature tensor in the
spinor-representation,
\begin{eqnarray}
[\nabla_M,\nabla_N]\ese\mF_{MN}= F_{MN}+R_{MN},\nonumber \\
F_{MN}\ese \pa_M A_N-\pa_N A_M+[A_M,A_N ]=F^a_{MN}T_a,\label{dirop11}\\
R_{MN}\ese \pa_M \Omega_N-\pa_N \Omega_M+[\Omega_M,\Omega_N]
=\q R_{MNAB}\Gamma^{AB},  \nonumber
\end{eqnarray}
where the Riemann curvature tensor is obtained
from the connection via
\eqnl{
R_{MNAB}=\pa_M\Omega_{NAB}-\pa_N\Omega_{MAB}
+\Omega_{MA}^{\hskip5mm C}\,\Omega_{NCB}
-\Omega_{NA}^{\hskip 5mm C}\,\Omega_{MCB},}{dirop15}
we find the squared Dirac operator or super-Hamiltonian
\eqnl{
-\di^2=H = Q_1^2=-G^{MN}
\nabla_M \nabla_N -\h \Gamma^{AB}\mF_{AB}.}{dirop16}
Here we have used the components of $\mF_{MN}$
with respect to an orthonormal vielbein,
\eqnn{
\mF_{AB}=\EAl\EBl\mF_{MN}=[\nabla_A,\nabla_B].}
Note that the two covariant derivatives $\nabla_M\nabla_N$
in (\ref{dirop16}) act on different types of fields. The
derivative on the
right acts on spinors and is given in \refs{dirop5}, whereas
the derivative on the left acts on spinors with a coordinate index and
hence contains an additional term proportional to
the Christoffel symbols,
\eqnl{
\nabla_M \psi_N =\pa_M \psi_N + \Omega_M \psi_N -
\Gamma^P_{MN} \psi_P + A_M \psi_N
\;. }{dirop18}

\subsection{Extended Supersymmetries}

In this section we show that for particular
background fields the chiral supersymmetry
can be extended to finer,
particle-number conserving supersymmetries.
The existence of a single conserved number operator
is equivalent to the existence of
a covariantly conserved  complex structure.
This way one finds that $\mN\es 2$ is only
possible if space admits a K\"ahler structure,
and $\mN\es 4$, if it admits a
hyper-K\"ahler structure. Analogous
conditions are derived for the background
gauge field.

\subsubsection{Square Roots of $H=-\di^2$}

In this subsection we characterize a class
of first order differential operators which
square to $H\es -\di^2$. Our ansatz is motivated
by previous results in
\cite{Carter:1979fe, Cooper:1988az, vanHolten:1999hh, Cotaescu:2003cs}
and the simple observation that both the free
Dirac operator $\fdi$ on flat space and
\eqnn{
\In{M}{N}\Gamma^N \pa_M}
have the same square for any orthogonal matrix $I$.
Thus, we are led to the following ansatz
for the supercharge in a gravitational
and gauge field background,
\eqnl{
Q(I)=\ui\,\In{M}{N}\Gamma^N\nabla_M\equiv
\ui(I\Gamma)^M\nabla_M,}{ext1}
where $I$ is a real tensor field with components $\In{M}{N}$.

This construction is close in spirit to the one
presented in \cite{Rittenberg:1983}. The algebraic
approach there is applied to the particular situation of
a Dirac operator, and this will allow us to interpret all 
quantities in \cite{Rittenberg:1983} as geometrical ones, 
like connections, curvature etc.

To derive the conditions on $I$ and the background such that
$Q(I)^2\es H$, we first consider the anticommutator
of two operators with different $I$,
\eqngrl{
\{Q(I),Q(J)\}\ese -\h(I J^T+J I^T)^{MN}\{\nabla_M,\nabla_N\}
-\h\Gamma^{MN}\left(I^T\mF J+J^T\mF I\right)_{MN}}
{&&-\left\{(I\Gamma)^P\nabla_P (J\Gamma)^Q
+(J\Gamma)^P\nabla_P (I\Gamma)^Q\right\}\nabla_Q,}{ext2}
where, for example
\eqnn{
(IJ^T)^{MN}=\In{M}{P}J^{NP}\mtxt{and}
(I^T\mF J)_{MN}
\es\In{P}{M}\mF_{PQ}\Jn{Q}{N}.}
After setting $I\es J$ we see that
$Q(I)$ squares to the Hamiltonian $H$ in \refs{dirop16} if and
only if the following three conditions are satisfied,
\begin{eqnarray}
G^{MN}\ese (II^T)^{MN},\label{ext4}\\
\mF_{MN}\ese (I^T\mF I)_{MN},\label{ext5}\\
0\ese \nabla_M \In{P}{Q}.\label{ext6}
\end{eqnarray}
By interpreting the $\In{M}{N}$ as components of
a map $I$ between sections of the tangent bundle, the condition
\refs{ext4} just means that $I$ is an isometry,
\eqnn{
G(IX,IY)=G(X,Y).}
In view of our remarks above it should not be surprising
that the components $\In{A}{B}$ with respect to an
orthonormal vielbein form an orthogonal matrix.

The condition \refs{ext6} means, that the tensor field
$I$ must be covariantly constant. With $R_{MNAB}=R_{ABMN}=
-R_{NMAB}$ the corresponding integrability conditions read
\eqngr{
0=\In{R}{M}[\nabla_A,\nabla_B]I_{RN}\ese
\In{R}{M}R_{RSAB}\In{S}{N}-
R_{MNAB}}
{\mtxt{or}
R_{MN}\ese(I^TRI)_{MN},}
and \refs{ext5} simplifies to the same condition
with $\mF_{MN}$ replaced by the gauge
field strength $F_{MN}$. Thus we end up with
the following

\vskip2mm
\textbf{Lemma:} \emph{The charge}
\eqnl{
Q(I)= \ui\,\In{M}{N}\Gamma^N\nabla_M,\qquad
\In{M}{N}(x)\in\R,}{ext10}
\emph{is Hermitian and squares to $H$
in \refs{dirop16} if and only if the following
conditions hold:}
\eqnl{
\nabla I=0,\quad II^T=\id \quad \textit{and} \quad[I,F]=0.}{ext11}

\vskip1mm
The hermiticity follows from $\nabla I\es 0$,
which in turn implies that the $\In{M}{N}\Gamma^N$
commute with the covariant derivative. Because of
the second condition in \refs{ext11} the last one is
equivalent to \refs{ext5} with
$\mF$ replaced by $F$.

A trivial solution is of course $I\es\id$ in which
case $Q$ becomes the Dirac operator itself. Let us now assume
that there is a second square root $Q(I)$
anticommuting with the Dirac operator $Q(\id)$.
With $\nabla I\es 0$ and \refs{ext2} these
two charges anticommute if
\eqnn{
\{Q(\id),Q(I)\}
=-\h(I^{MN}+I^{NM})\{\nabla_M,\nabla_N\}
-\frac{1}{2}\Gamma^{MN}\left(I^T\mF +\mF I\right)_{MN}
\stackrel{!}{=}0,}
and this shows that the map $I$
must be antisymmetric. Because of \refs{ext4}
it squares to $-\id$.
Hence it defines an \emph{almost complex structure.}
Since it is covariantly constant, the
manifold is K\"ahler with
complex structure $I$.
Thus we have shown that $H$ admits
two supersymmetries generated by
$Q(\id)$ and $Q(I)$ if the manifold is K\"ahler
with complex structure $I$ and if the gauge field
strength $F$ commutes with this structure.

Now we are ready to generalize to $\mN$ supercharges
\eqnn{
Q(\id) \mtxt{and} Q(I_i),\qquad i=1,\dots,\mN-1.}
From our general result \refs{ext2} we conclude

\vskip2mm
\textbf{Lemma:} \emph{The $\mN$ charges}
\eqnl{
Q(\id)=\ui\di \quad \textit{and} \quad
Q(I_i)= \ui\,\Ini{M}{N}\Gamma^N\nabla_M,
\quad i=1,\dots,\mN-1}{ext15}
\emph{are Hermitian and generate an extended superalgebra
\refs{sqm1}, if and only if}
\begin{eqnarray}
\{I_i,I_j\}=-2\delta_{ij}\id_D, && I_i^T=-I_i, \label{ext16}\\
\nabla I_i=0, && [I_i,F]=0.\label{ext17}
\end{eqnarray}

\vskip1mm
The covariantly conserved complex structures form
a $D$-dimensional real representation
of the Euclidean Clifford algebra with
$\mN\!-\!1$ gamma-matrices. We call a representation
irreducible, if only $\mathbbm{1}$ commutes with all 
those matrices. Irreducible representations
are known to exist for
\eqnl{
\begin{array}{rccc}
\mN-1 = & 8n,  & 6+8n,        & 7+8n,\\
D     = & 16^n,& 8\cdot 16^n, & 8\cdot 16^n,\\ 
\end{array}}{ext18}
with $n\in\N_0$. In these cases, only trivial gauge fields on flat space
are possible. We further observe that, if $\{I_1,\dots,I_k,F\}$
satisfy the conditions in the above lemma, then also
$\{I_1,\dots,I_k,I_{k+1}=I_1\cdots I_k,F\}$ do,
provided
\eqnn{k=2+4n.}
It follows, for example, that the superalgebra
with $7$ supercharges $\{Q(\id),Q(I_i)\}$ can always
be extended to a superalgebra with $8$ supercharges.
In addition, since for $\mN\ms  1=8n$ the
Euclidean gamma-matrices may be chosen
real and chiral, one can construct the corresponding
complex structures out of the complex structures $\tilde I_i$
of the $\mN=8n$ supersymmetry,
\eqnn{
I_i=\begin{pmatrix} 0 & \tilde I_i \\ \tilde I_i & 0 \end{pmatrix} \mtxt{and}
I_{8n}=\begin{pmatrix} 0 & \id \\ -\id & 0 \end{pmatrix}.}
For irreducible $I_i$ one
can construct systems with
\eqnl{\mN=8n \quad \textrm{and} \quad \mN=8n+1}{ext20}
independent real supercharges in this way.
Note, however, that there may exist $D$-dimensional
matrices $I_i$ which do not generate all of GL($D$)
and hence do not belong to a real irreducible representation of
the Clifford algebra. These are the most interesting
cases since they admit nontrivial background fields
commuting with all $I_i$, as required in our
lemma above. Below we will discuss such systems with
$\mN=4+8n$.

\subsubsection{$\mN\es 2$ and Particle-Number Operator}

On any K\"ahler manifold the Dirac operator
admits an extended $\mN\es 2$ supersymmetry
if the field strength commutes with the
complex structure. With respect to a
suitably chosen orthonormal frame the structure has
the form $(\In{A}{B})=\ui \sigma_2\otimes\id_n$.
The charge $Q(I)$ on a K\"ahler manifold with
complex structure $I$ squares to $H$ and commutes
with the Dirac operator if and only if
\eqnn{
[I,F]=0\mtxt{or}
(F_{AB})=\begin{pmatrix} U & V \\ - V & U \end{pmatrix},\quad
U^T=-U,\;V^T=V.}
The complex nilpotent charge in \refs{sqm10}
takes the simple form
\eqnl{
Q=\h Q(\id)+\ih Q(I)
=\ui\psiA\nabla_A}{neo0}
with operators
\eqnl{
\psiA=\PAu\Gamma^B,\qquad \PAu
=\h(\id+\ui I)^A_{\;\;B}.}{neo1}
$P$ projects onto the $n$-dimensional $I$-eigenspace
corresponding to the eigenvalue $-\ui$, its complex
conjugate $\bar P$ onto the $n$-dimensional
eigenspace $+\ui$. The two eigenspaces
are complementary and orthogonal, $P+\bar P\es\id$
and $P\bar P\es 0$. The $\psiA$ and their adjoints form a
fermionic algebra,
\eqnl{
\{\psiA,\psiB\}=\{\psidA,\psidB\}=0\mtxt{and}
\{\psiA,\psidB\}=2P^{AB}.}{neo3}
At this point it is natural to introduce the
\emph{number operator}
\eqnl{
N=\h\psi^\dagger_A\psiA=\q\left(D+\ui I_{AB}\Gamma^{AB}\right)
,}{neo4}
whose eigenvalues are lowered and raised
by $\psiA$ and $\psidA$, respectively,
\eqngrl{
[N,\psidA]\ese \;\PAu \psidB=\;\;\psidA,}
{\ [N,\psiA]\ese -\PAu\psiB=-\psiA.}{neo6}
Since only $n\es$ rank$\,P$ of the $2n$
creation operators are linearly independent
we have inserted a factor $\h$ in the definition
of the number operator $N$ in \refs{neo4}.
This operator commutes with the
covariant derivative, because $\nabla I=0$
is equivalent to
\eqnl{
[\nabla_M,N]=\pa_M N+[\Omega_M,N]=0,}{neo8}
and therefore $Q$ decreases $N$
by one, while its adjoint $Q^\dagger$
increases it by one,
\eqnl{
[N,Q]=-Q\mtxt{and} [N,Q^\dagger]=Q^\dagger.}{neo9}
The corresponding real supercharges are given
by
\eqngrl{
Q(\id)\ese Q+Q^\dagger=\ui\di,}
{Q(I)\ese \ui(Q^\dagger-Q)=\ui [N,\ui \di].}{neo10}
Finally, we observe that the Hermitian matrix
\eqnn{
\Sigma=N-\q D=\iq I_{AB}\Gamma^{AB}\in\hbox{spin}(D)}
generates a U$(1)$ subgroup of Spin$(D)$. This is
the R-symmetry of the superalgebra,
\eqnn{
\begin{pmatrix} Q (\id) \\ Q(I) \end{pmatrix} \longrightarrow
\begin{pmatrix} \cos\al & \sin\al \\ -\sin\al & \cos\al \end{pmatrix}
\begin{pmatrix} Q(\id) \\ Q(I) \end{pmatrix}.}
Next, we introduce the \emph{Clifford vacuum} $\va$,
which is annihilated by all annihilation
operators $\psiA$ and hence
has particle number $N\es 0$.
The corresponding Clifford space $\mathcal{C}$
is the Fock space built over this vacuum state.
Since only $n$ creation operators
are linearly independent, we obtain the
following grading of the Clifford space,
\eqnl{
\mathcal{C}= \mathcal{C}_0 \oplus \mathcal{C}_1 \oplus \ldots
\oplus \mathcal{C}_n, \quad \dim \mathcal{C}_p =\binom{n}{p},}{neo12}
with subspaces labelled by their particle number,
\eqnl{
N\big\vert_{\mathcal{C}_p}=p\cdot\id.}{neo14}
In particular, the one-dimensional subspace $\mathcal{C}_0$
is spanned by $\va$ and the $n$-dimensional
subspace $\mathcal{C}_1$ by the linearly dependent
states $\psi_A^\dagger\va$.
Along with the Clifford space the Hilbert space of
all square integrable spinor fields decomposes as
\eqnl{
\mH = \mH_0 \oplus \mH_1 \oplus \ldots \oplus \mH_n \quad
\mtxt{with} \quad N|_{\mH_p} = p \cdot \id. }{neo15}
Similar to the standard Fock space construction, cf. e.g.
\cite{Green:1987mn}, the number operator $N$ in \refs{neo4}
commutes with the Hamiltonian $H$ even in curved space and
in the presence of gauge fields. Thus, $H$ leaves
$\mH_p$ invariant. The nilpotent charge $Q$ maps $\mH_p$
into $\mH_{p-1}$ and its adjoint $Q^\dagger$ maps $\mH_p$
into $\mH_{p+1}$.

The raising and lowering operators $\psidA$ and $\psiA$
are linear combinations of $\Gamma^A$ and therefore
anticommute with $\Gamma$ in \refs{dirop7}.
Hence they map left- into right-handed spinors and vice
versa. Since $\Gamma\va$ is annihilated by all $\psiA$,
\eqnn{
\psiA \big(\Gamma \va \big) = - \Gamma \psiA \va = 0,}
and since the Clifford vacuum $\va$ is unique, we
conclude that $\va$ has definite chirality. It follows
that all states with even $N$ have
the same chirality as $\va$, and all states with odd $N$ have
opposite chirality,
\eqnl{
\Gamma = \pm (-)^N.}{neo11}

\subsubsection{$\mN=3$ and $\mN=4$ Superalgebras}

If $\{I_1,I_2\}$ satisfy the conditions
(\ref{ext16},\ref{ext17}), then $\{I_1,I_2,I_3=\pm I_1I_2\}$ do
so as well. For this reason $\mN\es 3$ supersymmetry implies
automatically $\mN\es 4$ supersymmetry. Hence, it suffices to consider
systems with $4$ supercharges. This should be compared to the
results in \cite{Hull:1999ng}, where systems with
$\mN=3$ but $\mN\neq 4$ are possible, the reason for this
being that in \cite{Hull:1999ng} a more general algebra
than \refs{sqm1} has been studied.

The dimension
of the matrices $I_i$ (which equals the
dimension of the manifold) must be a multiple
of $4$, $D\es 4n$. In this section we choose
the selfdual or anti-selfdual matrices,
\eqngrl{
\textrm{SD:}\qquad
\tilde I_1=\ui\sigma_0\otimes\sigma_2,\quad&
\tilde I_2=\ui\sigma_2\otimes\sigma_3,\quad&
\tilde I_3=\ui\sigma_2\otimes\sigma_1=- \tilde I_1\tilde I_2,}
{\textrm{ASD:}\qquad\tilde I_1=\ui\sigma_3\otimes \sigma_2,\quad&
\tilde I_2=\ui\sigma_2\otimes \sigma_0,\quad&
\tilde I_3=\ui\sigma_1\otimes \sigma_2=\tilde I_1\tilde I_2,}{nez1}
and define $I_i = \tilde I_i \otimes \id_n$.
They generate two commuting $so(3)$ subalgebras
of $so(4n)$. The conditions
(\ref{ext16},\ref{ext17}) imply that the
curvature tensor $(R_{AB})$ and gauge field strength
$(F_{AB})$ commute with all three $I_i$. For example,
in $4$ dimensions both must be \emph{selfdual} or
\emph{anti-selfdual}. A $4$-dimensional manifold
with (anti-)selfdual curvature is hyper-K\"ahler.
More generally, a $4n$-dimensional manifold
is hyper-K\"ahler if it admits three covariantly
constant and anticommuting complex structures.
We see, that $-\di^2$ admits $4$ supersymmetries
if and only if the underlying space ${\cal M}$ is hyper-K\"ahler
and the gauge field strength commutes with the
three complex structures.

For each complex structure $I_i$ there exists an
associated number operator
\eqnl{
N_i = N(I_i)=\q D+\Sigma(I_i),\qquad
\Sigma(I)=\iq I_{AB}\Gamma^{AB},}{nez5}
and the $4$ real supercharges take the form
\eqnl{
Q(\id)=\ui \di\mtxt{and}
Q(I_i)=\ui [N_i,\ui \di].}{nez4}
However, the 3 number operators do not commute, because
\eqnl{
[\Sigma(I_i),\Sigma(I_j)]=
\ui\Sigma\left([I_i,I_j]\right),}{nez6}
and the antisymmetric matrices $I_i$, together
with $\id_{4n}$, form a
$4n$-dimensional real representation of the
\emph{non-commutative} quaternionic algebra,
\eqnl{
I_iI_j=-\delta_{ij}\id_{4n}\pm \epsilon_{ijk}I_k.}{nez7}
The three matrices $\Sigma(I_i)$ generate an SO$(3)$-subgroup
of Spin$(4n)$ which rotates the real supercharges.
This is proven with the help of the simple
identities
\eqnn{
\ui [\Sigma(I),Q(\id)]=Q(I)\mtxt{and}
\ui [\Sigma(I),Q(J)]=Q(JI).} 
Now it follows at once, that the selfdual
(anti-selfdual) SO$(3)$-subgroup of the
SO$(4)$ R-symmetry is implemented
by the exponentiated action of the $\Sigma(I_i)$, 
\eqngr{
&&U(\vec\al)Q_m U^{-1}(\vec\al)=R_{mn}Q_n,\qquad\hbox{where}}
{&&U(\vec\al)=\exp\left(\ui\Sigma(\al_i I_i)\right)
,\quad
R(\vec\al)=\exp(\al_i \tilde I_i).}The $\tilde I_i$ are the $4$-dimensional
selfdual (anti-selfdual) matrices in \refs{nez1},
and $I_i\es\tilde I_i\otimes \id_n$ are
$4n$-dimensional complex structures with respect
to a suitable orthonormal base. The $Q_m$ are the
four real supercharges,
\eqnn{
\{Q_0,Q_1,Q_2,Q_3\}\equiv
\{Q(\id),Q(I_1),Q(I_2),Q(I_3)\}.}
Let us remark, that other choices for the complex
structures than those obtained from \refs{nez1}
are possible.

\subsubsection{$\mN=7$ and $\mN=8$ Superalgebras}

According to \refs{ext18} we can find
$6$ or $7$ real and antisymmetric matrices
$I_i$, for example the $8$-dimensional (irreducible) matrices
\eqngrrl{
&& \tilde I_1=\ui\sigma_1\otimes\sigma_0\otimes \sigma_2,\quad
\tilde I_3=\ui\sigma_2\otimes\sigma_1\otimes\sigma_0,\quad
\tilde I_5=\ui\sigma_0\otimes\sigma_2\otimes\sigma_1,}
{&&\tilde I_2=\ui\sigma_3\otimes\sigma_0\otimes\sigma_2,\quad
\tilde I_4=\ui\sigma_2\otimes\sigma_3\otimes \sigma_0,\quad
\tilde I_6=\ui\sigma_0\otimes\sigma_2\otimes\sigma_3,}
{&& \tilde I_7 = \tilde I_1 \tilde I_2 \tilde I_3 \tilde I_4 \tilde I_5
\tilde I_6 = -\ui\sigma_2\otimes\sigma_2\otimes\sigma_2,}{nef1}
tensored with $\id_n$. Thus we can satisfy \refs{ext16}
in $8n$ dimensions and a $\mN\es 7$ superalgebra
can always be extended to a $\mN\es 8$ superalgebra,
since if $\{I_1,\dots,I_6,F\}$ satisfy the
conditions (\ref{ext16},\ref{ext17}) then
$\{I_1,\dots,I_6,I_7= I_1\cdots I_6,F\}$
do so as well.

In $8$ dimensions there is no non-trivial solution to
\eqnn{
[I_i,\mF]=0,\quad i=1,\dots,7,}
since the only matrix commuting with all
$I_i$ in \refs{nef1} is the identity matrix.
Hence the manifold must be flat and the gauge
field strength must vanish. In $8$
dimensions, only the free Dirac operator admits an
$\mN\es 8$ supersymmetry. However, in $8n$ dimensions
with $n\es 2,3,\dots,$ there are nontrivial solutions
to the constraints in (\ref{ext16},\ref{ext17}).
For example, every field strength $(F_{AB})=\id_8\otimes
\tilde F$ with antisymmetric $\tilde F$
commutes with the $I_i$ listed in \refs{nef1}.
In the case of extended supersymmetry one
can define a set $\{N_{01}, N_{23}, N_{45}, \ldots \}$
of particle-number operators that commute
with each other and with the Hamiltonian. Here the
$N_{ij}$ are defined as
\eqnl{N_{ij} = N(I_i I_j) = \frac{1}{4}D + \Sigma(I_i I_j),
\quad \textrm{where} \quad I_0 = \id.}{nef2}

\section{Superpotentials on K\"ahler Manifolds}
\label{superpotential}

The super-Hamiltonian $-\di^2$ admits an
extended supersymmetry if it commutes
with a number operator or if the
complex supercharge is nilpotent and decreases
the particle number by one.
Then the manifold is K\"ahler and the complex
structure commutes with the gauge field strength.
Now we shall see that this in turn is the condition
for the existence of a superpotential $g$ from
which the spin connection and gauge potential
can be derived.

K\"ahler manifolds of real dimension $D=2n$ are particular complex manifolds
and we may introduce complex coordinates
$(z^\mu,\bar z^{\bar \mu})$ with $\mu,\bar\mu=1,\dots,n$
\cite{Candelas:1987is}.
The real and complex coordinate differentials
are related as follows
\eqngrl{
\ud z^\mu=\frac{\pa z^\mu}{\pa x^M} \ud x^M \equiv \fMl \ud x^M, && \quad 
\ud \bar z^{\bar\mu}=\frac{\pa \bar z^{\bar \mu}}{\pa x^M} \ud x^M
\equiv\fbMl \ud x^M,}
{\pa_\mu=\frac{\pa x^M}{\pa z^\mu}\pa_M\equiv \fMu \pa_M, && \quad 
\pa_{\bar\mu}=\frac{\pa x^M}{\pa\bar z^{\bar\mu}}\pa_M\equiv
\fbMu\pa_M.}{com1}
The integrability conditions for the $\ud z^\mu$
to be differentials of complex coordinate functions
$z^\mu$ are automatically satisfied on a K\"ahler manifold.

The $f^\mu$ and $f_\mu$ are left and right eigenvectors
of the complex structure,
\eqnl{
\fMl\IMu=-\ui \fNl\mtxt{and}
\IMu\fNu=-\ui \fMu,\qquad \mu=1,\dots,n.}{com2}
Since $\IMu$ is antisymmetric with
respect to the scalar product $(A,B)\es A_MG^{MN}B_N$,
the eigenvectors with different eigenvalues
are orthogonal in the following sense,
\eqnl{
G^{MN}\fMl f^\nu_{\hskip1.6mm N}=G_{MN}\fMu f^N_{\hskip2.5mm\nu}=0.}{com3}
Identity and complex structure possess the spectral decompositions
\begin{eqnarray}
\delta^M_{\hskip 2mm N}=\fMu\fNl+\fbMu\fbNl, \label{com4}\\
\ui\IMu =\fMu\fNl-\fbMu\fbNl,\label{com5}
\end{eqnarray}
and the relations  $\pa z^\mu/\pa z^\nu\es\delta^\mu_\nu$
and $\pa z^\mu/\pa \bar z^{\bar \nu}=0$ translate into
\eqnl{
\fMl f^M_{\hskip2.5mm\nu}=\delta^\mu_{\hskip1.5mm \nu}\mtxt{and}
\fMl f^M_{\hskip2.5mm \bar\nu}=0.}{com6}
With \refs{com3} the line element takes the form
\eqnl{
\ud s^2=G_{MN} \ud x^M \ud x^N = 2h_{\mu\bar\nu} \ud z^\mu \ud\bar z^{\bar\nu}
,\quad
h_{\mu\bar\nu}=h_{\bar\nu \mu}=
G_{MN}\fMu\,f^N_{\hskip2.5mm \bar\nu},}{com8}
where the $h_{\mu\bar\nu}$ are derived from a real K\"ahler
potential $K$ as follows,
\eqnl{
h_{\mu\bar \nu}=\frac{\pa^2 K}{\pa z^\mu \pa \bar z^{\bar\nu}}
\equiv \pa_\mu \pa_{\bar\nu}K.}{com10}
Covariant and exterior derivatives split into
holomorphic and antiholomorphic pieces, 
\eqngrl{
\nabla\ese \ud z^\mu\nabla_\mu+\ud\bar z^{\bar\mu} \nabla_{\bar\mu},}
{\ud\ese \ud z^\mu \pa_\mu+\ud\bar z^{\bar\mu}
\pa_{\bar \mu}=\pa+\bar\pa,}{com12}
and the only non-vanishing components of
the Christoffel symbols are 
\begin{eqnarray}
\Gamma^\rho_{\mu\nu}\ese
h^{\rho\bar\sigma}\pa_\mu h_{\bar\sigma\nu}=
h^{\rho\bar\sigma}\pa_{\bar\sigma\mu\nu}K,\label{com13a}\\
\Gamma^{\bar \rho}_{\bar\mu\bar\nu}
\ese h^{\bar\rho\sigma}\pa_{\bar\mu}h_{\sigma\bar\nu}=
h^{\bar\rho\sigma}\pa_{\sigma\bar\mu\bar\nu}K.\label{com13b}
\end{eqnarray}
Along with the derivatives the forms split
into holomorphic and antiholomorphic parts.
For example, the first Chern class $c_1 = (2\pi\ui)^{-1} h_{\mu\bar\nu}
\ud z^\mu \ud \bar z^{\bar\nu}$
is a $(1,1)$-form and the gauge potential
$A= A_\mu \ud z^\mu + A_{\bar \mu}\ud\bar z^{\bar\mu}$
a sum of a $(1,0)$- and a $(0,1)$-form.
With the help of \refs{com13a} the covariant derivative
of a $(1,0)$-vector field can be written as
\eqngrrl{
\nabla_\mu (B^\nu\pa_\nu)\ese
\big(\pa_\mu B^\rho+\Gamma_{\mu\nu}^\rho B^\nu\big)
\pa_\rho}
{\ese
\big(\pa_\mu B^\rho+h^{\rho\bar\sigma}
(\pa_{\mu}h_{\bar\sigma\nu}) B^\nu\big)\pa_\rho}
{\ese
h^{\rho\bar\sigma}\pa_\mu (h_{\bar\sigma\nu} B^\nu)\pa_\rho.}{com14}
Let us introduce complex vielbeins $e_\alpha \es e^\mu_\alpha \partial_\mu$
and $e^\alpha \es e^\alpha_\mu \ud z^\mu$, such that $h_{\bar\mu\nu} = \frac{1}{2} \delta_{\bar \alpha \beta} e_{\bar\mu}^{\bar\alpha} e_\nu^\beta$. The components of the complex connection can
be related to the metric $h_{\bar\mu\nu}$ and the vielbeins with the
help of Leibniz' rule and \refs{com13a} as follows,
\eqngr{
\om_{\mu\al}^\beta e_\beta \!&\equiv&\! \nabla_\mu e_\al
=\nabla_\mu (e^\nu_\al \pa_\nu)=(\pa_\mu e^\nu_\al) \pa_\nu
+e^\nu_\al \Gamma_{\mu\nu}^\rho\pa_\rho}
{\ese (\pa_\mu e^\nu_\al) \pa_\nu
+e^\nu_\al h^{\rho\bar\sigma}\pa_\mu (h_{\bar\sigma \nu})\pa_\rho
=h^{\rho\bar\sigma}\pa_\mu (e^\nu_\al h_{\bar\sigma\nu})\pa_\rho
=e_\rho^\beta h^{\rho\bar\sigma}\pa_\mu (e^\nu_\al h_{\bar\sigma\nu})
e_\beta.}
Comparing the coefficients of $e_\beta$ yields the
connection coefficients $\om_{\mu\al}^\beta$.
The remaining coefficients are obtained the same
way, and one finds altogether
\eqngrl{
\om_{\mu\al}^\beta=e^{\beta\bar\sigma}\pa_\mu e_{\bar\sigma\al},&&
\om_{\mu\bar\al}^{\bar\beta}=e^{\bar\beta}_{\bar\sigma}
\pa_\mu e^{\bar\sigma}_{\bar \al},}
{\om_{\bar\mu\bar\al}^{\bar\beta}
=e^{\bar\beta\sigma}\bar\pa_{\bar\mu} e_{\sigma\bar\al}
,&&
\om_{\bar\mu \al}^{\beta}=e^\beta_\sigma
\bar\pa_{\bar\mu} e^\sigma_\al,}{com16}
where, for example, $e^{\beta\bar\sigma} \es
h^{\bar \sigma \rho} e_{\rho}^{\beta}$.

Now we are ready to rewrite the Dirac operator
in complex coordinates.
For that we insert the completeness relation \refs{com4}
in $\ui\di= \ui \Gamma^N\delta^M_{\hskip 2mm N}\nabla_M$ and
obtain
\eqnl{
\ui\di=Q+Q^\dagger\equiv 2\ui \psi^\mu\nabla_\mu
+2\ui \psi^{\dagger \bar \mu}\nabla_{\bar\mu},}{com20}
where we are led to the independent fermionic
raising and lowering operators,
\eqnl{
\psi^\mu=\h\fMl\Gamma^M,\quad
\psi^{\dagger\bar\mu}=\h\fbMl\Gamma^M,}{com21}
and the complex covariant derivatives
\eqnl{
\nabla_\mu=\fMu\nabla_M,\quad
\nabla_{\bar\mu}=\fbMu\nabla_M.}{com22}
Of course, the supercharge $Q$ in \refs{com20}
is just the charge in \refs{neo0} rewritten
in complex coordinates. Contrary to the
annihilation operators $\psiA$, the
fermionic operators $\psi^\mu$ are independent. 
They fulfill the anticommutation relations
\eqnl{
\{\psi^\mu,\psi^\nu\} = \{\psi^{\dagger\bar\mu},\psi^{\dagger\bar\nu}\}
\stackrel{\refs{com3}}{=}0, \qquad
\{\psi^\mu,\psi^{\dagger\bar\nu}\} = \h h^{\mu\bar\nu},}{com23}
where $h^{\mu\bar\nu}=\fMl f^{\bar \nu M}$ is
the inverse of $h_{\mu\bar\nu}$ in \refs{com8}. This can be
seen as follows,
\eqngr{
h^{\bar\mu\sigma} h_{\sigma\bar\nu}
\ese f^{\bar\mu M}f^\sigma_{\hskip 1.6mm M}
\cdot f^N_{\hskip 2mm \sigma}f_{N\bar \nu}
\stackrel{\refs{com3}}{=}
f^{\bar\mu M}\left(f^N_{\hskip 2mm \sigma} f^\sigma_{\hskip 1.6mm M}+
f^N_{\hskip 2mm \bar\sigma} f^{\bar\sigma}_{\hskip 1.6mm M}\right)f_{N\bar \nu}}
{\!&\stackrel{\refs{com4}}{=}&\! f^{\bar\mu M}f_{M\bar\nu}
\stackrel{\refs{com6}}{=}
\delta^{\bar\mu}_{\hskip1.6mm \bar\nu}\,. }
The operators $\psi^\mu$ lower
the value of the Hermitian number operator
\eqnl{
N=2h_{\bar\mu \nu}\psi^{\dagger\bar\mu}\psi^\nu}{com25}
by one, while the $\psi^{\dagger\bar\mu}$
raise it by one. The proof is simple,
\eqnn{
[N,\psi^\sigma]=2h_{\bar\mu\nu}[\psi^{\dagger\bar\mu}\psi^\nu,\psi^\sigma]
=-2h_{\bar\mu\nu}\{\psi^{\dagger\bar\mu},\psi^\sigma\}\psi^\nu
=-h_{\bar\mu\nu}h^{\bar\mu\sigma}\psi^\nu=-\psi^\sigma.}
With (\ref{com4},\ref{com5}) the
fermionic operators in \refs{neo1} and
\refs{com21} are related as follows,
\eqngrl{
\psi^M\ese \h(\id+\ui I)^M_{\hskip 2mm N}\Gamma^N
=2\fMu\psi^\mu,}
{\psi^{\dagger M}\ese\h(\id-\ui I)^M_{\hskip 2mm N}\Gamma^N
=2\fbMu \psi^{\dagger\bar\mu},}{com26}
and we conclude that the number operators
in \refs{neo4} and \refs{com25} are indeed equal,
\eqnn{
\h\psi^{\dagger M}\psi_M
=2G_{MN}\fbMu f^N_{\hskip2.5mm \nu}
\psi^{\dagger\bar\mu}\psi^\nu=2h_{\bar\mu\nu}\psi^{\dagger\bar\mu}
\psi^\nu.}
Now we are ready to prove that in cases where
$\di$ admits an extended supersymmetry there
exists a superpotential for the spin and
gauge connections. Indeed, if spacetime
is K\"ahler and the gauge field strength
commutes with the complex structure,
\eqnl{
F_{MN}=(I^T F I)_{MN},}{com27}
then the complex covariant derivatives commute
\eqnl{ 
[\nabla_\mu,\nabla_\nu]=\mF_{\mu\nu}=
\fMu f^N_{\hskip 2.5mm \nu}\mF_{MN}=0.}{com28}
One just needs to insert \refs{com5}
and use \refs{com6}. Alternatively, one
may use $Q^2\es 0$ with the complex supercharge
in \refs{com20}.
Equation \refs{com28} is just the
integrability condition (cf. Yang's equation \cite{Yang:1977})
for the existence of a superpotential $g$ such that
the complex covariant derivative can be written as
\eqnl{
\nabla_\mu =g\pa_\mu g^{-1}=\pa_\mu+g\left(\pa_\mu g^{-1}\right)
=\pa_\mu+\om_\mu + A_\mu.}{com30}
This useful property is true for any
(possibly charged) tensor field on a K\"ahler manifold
provided \refs{com27} holds. If the K\"ahler manifold
admits a spin structure, as for example $\C P^n$ for odd $n$,
then \refs{com30} holds true for a (possibly charged) spinor field.

Of course, the superpotential $g$ depends on
the representation according to which the fields transform under
the gauge and Lorentz group. One of the more severe technical
problems in applications
is to obtain $g$ in the representation of interest. It consists
of two factors, $g=g_A g_\omega$. The first factor $g_A$
is the path-ordered integral of the gauge potential.
According to \refs{com16} and \refs{com30} the matrix $g_\omega$
in the vector representation is just the vielbein $e^{\beta\bar\sigma}$.
If one succeeds in rewriting this $g_\omega$ as the exponential
of a matrix, then the transition to any other representation
is straightforward: one contracts the matrix in the exponent with the
generators in the given representation. This will be done for the complex
projective spaces in the following section.

Now let us assume that we have found
the superpotential $g$. Then we can rewrite
the complex supercharge in \refs{com20} as follows,
\eqnl{
Q=2\ui \psi^\mu \nabla_\mu=
2\ui g Q_0 g^{-1},\quad
Q_0=\psi_0^\mu\pa_\mu,\quad \psi_0^\mu =g^{-1}\psi^\mu g.}{com32}
The annihilation operators $\psi^\mu$ are
covariantly constant,
\eqnl{
\nabla_\mu\psi^\nu=\pa_\mu\psi^\nu+\Gamma^\nu_{\mu\rho}\psi^\rho
+[\om_\mu,\psi^\nu] = 0,}{com33}
and this translates into the following
property of the conjugate operators,
\eqngr{
\pa_\mu\psi_0^\nu\ese g^{-1}\left(\pa_\mu \psi^\nu+
[\,g\pa_\mu g^{-1},\psi^\nu]\right)g
\stackrel{\refs{com30}}{=}
g^{-1}\left(\pa_\mu \psi^\nu+[\om_\mu,\psi^\nu]\right)g}
{\ese - \Gamma^{\nu}_{\mu\rho}\, g^{-1}\psi^\rho g
\stackrel{\refs{com13a}}{=}
-h^{\nu\bar\sigma}\left(\pa_\mu h_{\bar\sigma\rho}\right)
\psi_0^\rho.}
This implies the following simple equation,
\eqnl{
\pa_\mu \left(h_{\bar\sigma\rho}
\psi_0^\rho\right)=0,}{com34}
stating that the transformed annihilation
operators  $\psi_{0\bar\sigma}$ are antiholomorphic. Indeed,
one can show that they are even constant.

The relation \refs{com32} between the free supercharge $Q_0$
and the $g$-dependent supercharge $Q$ is the main result
of this section.
It can be used to determine zero modes of the Dirac operator. 
With \refs{neo10} we find
\eqnl{\ui \di \chi = 0 \quad \Longleftrightarrow \quad Q \chi = 0 , \; Q^\dagger \chi = 0.}{com35}
In sectors with particle number $N\es0$ or $N\es n$ one can easily solve
for all zero modes. For example, $Q^\dagger$ annihilates all states in the
sector with $N=n$, such that zero modes only need to satisfy $Q\chi = 0$.
Because of \refs{com32}, the general solution of this equation reads
\eqnl{\chi = \bar f(\bar z) g \psi^{\dagger \bar 1} \cdots \psi^{\dagger\bar n} \va,}{com36}
where $\bar f(\bar z)$ is some
antiholomorphic function. Of course, the number of \emph{normalizable}
solutions depends on the gauge and gravitational background fields
encoded in the superpotential $g$. With the help of the novel result
\refs{com36} we shall find the explicit form of the zero modes on $\C P^n$
in the following section.

\section{The Dirac Operator on $\C P^n$}

The ubiquitous two-dimensional $\C P^n$ models
possess remarkable similarities
with non-Abelian gauge theories
in 3+1 dimensions \cite{shifman:1984}.
They are frequently used as toy models displaying interesting
physics like fermion-number violation analogous to the electroweak
theory \cite{mottola:1989} or spin excitations
in quantum Hall systems \cite{rajaraman:2001}. Their
instanton solutions have been
studied in \cite{aguado:2002}, and their $\mN\es 2$
supersymmetric extensions describe integrable systems with
known scattering matrices.

It would be desirable to construct
manifestly supersymmetric extensions
of these models on a spatial lattice.
To this end we reconsider the Dirac operator
on the symmetric K\"ahler manifolds $\C P^n$.
We  shall calculate the superpotential $g$ in
\refs{com30} and the explicit zero modes
of the Dirac operator.

\subsection{Complex Projective Spaces}

First we briefly recall those properties of the
complex projective spaces $\C P^n$ which are
relevant for our purposes. The space $\C P^n$ consists
of complex lines in $\C^{n+1}$ intersecting the origin.
Its elements are identified with the following equivalence
classes of points $u=(u^0,\dots,u^n)\in \C^{n+1}\backslash \{0\}$,
\eqnn{
[u]=\{v=\lambda u\vert \lambda\in\C^*\},}
such that $\C P^n$ is identified with
$(\C^{n+1}\backslash \{0\})/\C^*$.
In each class there are elements with unit
norm, $\bar u\cdot u=\sum\bar u^j u^j\es 1$, and thus there
is an equivalent characterization
as a coset space of spheres, $\C P^n=S^{2n+1}/S^1$.
The $u$'s are homogeneous coordinates of $\C P^n$.
We define the $n\ps 1$ open sets
\eqnl{
U_k=\big\{u\in \C^{n+1}\vert u^k\neq 0\big\}
\subset \C^{n+1}\backslash \{0\},}{cpn1}
the classes of which cover the projective space.
The $n\ps 1$ maps
\eqnn{\varphi_k: \quad \C^n\longrightarrow [U_k], \quad
z\mapsto [z^1,\dots,1,\dots,z^n],}
where the $k^\textrm{th}$ coordinate is $1$, define a 
complex analytic structure.
The line element on $\C^{n+1}$, 
\eqnl{
\ud s^2 =\sum_{j=0}^n \ud u^j \ud\bar u^j = \ud u\cdot \ud\bar u,}{cpn3}
can be restricted to $S^{2n+1}/S^1$ and has
the following representation on the $k^\textrm{th}$ chart,
\eqnn{
\ud s^2 = \left(\frac{\pa u}{\pa z^\mu} \ud z^\mu
+\frac{\pa u}{\pa\bar z^{\bar\mu}}\ud\bar z^{\bar \mu}\right)
\cdot\left(\frac{\pa\bar u}{\pa z^\mu} \ud z^\mu
+\frac{\pa \bar u}{\pa\bar z^{\bar\mu}}\ud\bar z^{\bar \mu}\right).}
We shall use the (local) coordinates
\eqnl{
u=
\varphi_0(z)=
\frac{1}{\rho}\left(1,z\right)\in U_0,\mtxt{where}
\rho^2=1+\bar z\cdot z=1+r^2,}{cpn5}
for representatives with non-vanishing $u^0$.
With these coordinates the line element
takes the form 
\eqnl{
\ud s^2 =\frac{1}{\rho^2}\ud z\cdot \ud\bar z-\frac{1}{\rho^4}(\bar z\cdot \ud z)(z\cdot \ud\bar z),}
{cpn6}
and is derived from a K\"ahler potential $K\es\ln \rho^2$. This
concludes our summary of $\C P^n$.

To couple fermions to the gravitational background
field we must find a complex orthonormal vielbein,
$\ud s^2=e^\al \delta_{\al\bar\al} e^{\,\bar\al}$,
and write it as the exponential of a matrix. 
We obtained the following representation
for the vielbein of the Fubini-Study metric \refs{cpn6},
\eqngrl{
e^\al\ese e^\al_\mu \ud z^\mu
=\rho^{-1}\left(\mathbbm{P}_{\;\;\mu}^{\alpha} + \rho^{-1} \mathbbm{Q}_{\;\;\mu}^{\alpha} \right)\ud z^\mu,}
{e_\al\ese e_\al^\mu\pa_\mu
= \rho \left(\mathbbm{P}_{\;\;\alpha}^{\mu} + \rho \mathbbm{Q}_{\;\;\alpha}^{\mu} \right)\pa_\mu.} {cpn12}
Here, we have introduced the matrices
\eqnl{\mathbbm{P} = \id - \frac{\vecbf{z} \vecbf{z}^\dagger}{r^2} \mtxt{and} \mathbbm{Q} = \frac{\vecbf{z} \vecbf{z}^\dagger}{r^2}, \qquad \vecbf{z} = \left(z^1 \ldots z^n \right)^T.} {cpn12a}
They satisfy
\eqnl{\mathbbm{P}^2 = \mathbbm{P}, \qquad
\mathbbm{Q}^2 = \mathbbm{Q}, \qquad
\mathbbm{PQ} = \mathbbm{QP} = 0, \qquad
\mathbbm{P}^\dagger = \mathbbm{P}, \qquad
\mathbbm{Q}^\dagger = \mathbbm{Q},}{cpn12b}
and hence are orthogonal projection operators.
For the particular space $\C P^2$, the vielbeins are known, and
can be found in \cite{Eguchi:1980jx}. These known ones are related to
those in \refs{cpn12} by a local Lorentz transformation. We have not seen 
explicit formulae for the vielbeins for $n>2$ in the literature.
Expressing the vielbeins in terms of projection operators as
in \refs{cpn12} allows us to relate the superpotentials in
different representations. From \refs{com16} and \refs{cpn12} we obtain the connection (1,0)-form
\eqnn{
\om^\al_{\mu\beta} = - \frac{\bar z_\mu}{\rho^2} \left( \frac{1}{2}
\mathbbm{P}^\alpha_{\;\;\beta} + \mathbbm{Q}^\alpha_{\;\;\beta} \right) + \frac{1-\rho}{\rho r^2} \mathbbm{P}^\alpha_{\;\;\mu} \bar z_\beta.}

\subsection{Zero Modes of the Dirac Operator}

In this subsection we want to determine the zero modes of the
Dirac operator $\ui \di$ on $\C P^n$. We use the method proposed
at the end of Section \ref{superpotential}. Actually, only for odd
values of $n$ a spin bundle $S$ exists on $\C P^n$.
We can tensor $S$ with $L^{k/2}$, where $L$ is the generating
line bundle, and $k$ takes on even values. In the language of field
theory this means that we couple fermions to a U(1) gauge potential $A$.
For even values of $n$, there is no spin structure, so $S$ does not
exist globally. Similarly, for odd values of $k$, $L^{k/2}$ is
not defined globally. There is, however, the possibility to
define a generalized spin bundle $S_c$ which is the formal
tensor product of $S$ and $L^{k/2}$, $k$ odd \cite{Hawking:1978ab}.  
Again, in the language of field theory, we
couple fermions to a suitably chosen
U(1) gauge potential with half-integer instanton number. In both
cases, the gauge potential reads 
\eqnl{
A=\frac{k}{2}\bar u\cdot \ud u=
\frac{k}{4}(\pa-\bar\pa)K
= g_A \pa g^{-1}_A +g_A^{\dagger -1}\bar\pa g_A^\dagger,\quad
g_A= \ue^{-kK/4} = (1+r^2)^{-\frac{k}{4}},}{cpn8}
with corresponding field strength
\eqnl{
F=\ud A=(\pa+\bar \pa)A
=\frac{k}{2}\bar\pa\pa K.}{cpn9}
$g_A$ is the part of the superpontential $g$ that
gives rise to the gauge connection $A$. It remains
to determine the spin connection part $g_\omega$ of $g \equiv g_\omega g_A$.

When using (\ref{cpn12},\ref{cpn12a}), the equation \refs{com16}
can be written in matrix notation as $(\omega_\mu)^{\alpha}_{\;\;\beta}
= \left(S \partial_\mu S^{-1}\right)^{\alpha}_{\;\;\beta}$, where
\eqnl{S = \rho ( \mathbbm{P} + \rho \mathbbm{Q} ) \stackrel{\refs{cpn12b}}{=} \exp \left(  \mathbbm{P} \ln \rho+  \mathbbm{Q} \ln \rho^2 \right)= \exp \Big( (\id + \mathbbm{Q} )
\ln \rho \Big).}{cpn9a}
From the matrix form of $S$ in \refs{cpn9a} we read off the
superpotential $g_\omega$ in the \emph{spinor representation},
\eqnl{g_\omega = \exp \left( \frac{1}{4}(\delta_{\bar \alpha \beta} + \mathbbm{Q}_{\bar \alpha \beta}) \Gamma^{\bar \alpha \beta} \ln \rho \right),}{cpn9c}
where we have introduced
\eqnl{\Gamma^{\bar\alpha \beta} \equiv \frac{1}{2} [\Gamma^{\bar\alpha}, \Gamma^\beta] = 2 [\psi^{\dagger\bar\alpha}, \psi^\beta] , \qquad \Gamma^{\bar\alpha} = 2
\psi^{\dagger\bar\alpha}, \qquad \Gamma^\beta = 2 \psi^\beta.}{cpn9d}
Next, we study zero modes of $Q$ and $Q^\dagger$ in the gauge field
background \refs{cpn8}. In the sector of interest with $N\es n$, the superpotential
$g_\omega$ in the spinor representation simplifies as
\eqnl{g_\omega \big|_{N=n} = (1+r^2)^{\frac{n+1}{4}}, \quad \textrm{since} \quad
\Gamma^{\bar\alpha\beta}\big|_{N=n} = 2 \delta^{\bar\alpha\beta}.}{cpn16} 
All states in the $N=n$ sector are annihilated by $Q^\dagger$. Zero
modes $\chi$ satisfy in addition
\begin{eqnarray}
\label{cpn17}
0 = Q \chi = 2 \ui \psi^\mu \nabla_\mu \chi = 2 \ui \psi^\mu g \partial_\mu g^{-1} \chi, \qquad g = g_A g_\omega.
\end{eqnarray}
Using \refs{cpn8} and \refs{cpn16} we conclude that
\eqnl{\chi = g \bar f (\bar z)\psi^{\dagger \bar 1} \cdots \psi^{\dagger\bar n} \va
= (1+r^2)^{\frac{n+1-k}{4}} \bar f (\bar z)\psi^{\dagger \bar 1} \cdots \psi^{\dagger\bar n} \va,}{cpn17a}
 with some antiholomorphic function $\bar f$. Normalizability of
$\chi$ will put restrictions on the admissible functions
$\bar f$. Since the operators $\bar z^{\bar \mu} \partial_{\bar \mu}$ (no sum) commute with $\partial_{\mu}$ and with each other, we can diagonalize them
simultaneously on the kernel of $\partial_\mu$. Thus, we are let to the following most general ansatz
\begin{eqnarray}
\label{cpn18}
\bar f_m = (\bar z^{\bar 1})^{m_1} \cdots (\bar z^{\bar n})^{m_n}, \qquad \sum_{i=1}^n m_i = m.
\end{eqnarray}
There are $\binom{m+n-1}{n-1}$ independent functions
of this form. The solution $\chi$ in \refs{cpn17a} is square-integrable if and only if
\begin{eqnarray}
\label{cpn19}
\| \chi \|^2 & = & \int \ud \textrm{vol} \; (\det h) \; \chi^\dagger \chi \nonumber \\
& \stackrel{\refs{cpn17a}}{\propto} & \int \ud \Omega \int \ud r \; r^{2m+2n-1} (1+r^2)^{-\frac{n+k+1}{2}} < \infty, 
\end{eqnarray}
so normalizable zero modes in the $N=n$ sector exist for
\begin{eqnarray}
\label{cpn20}
m = 0,1,2, \ldots, q \equiv \frac{1}{2}(k-n-1).
\end{eqnarray}
Note, that $q$ is always integer-valued, since $k$ is
odd (even) if $n$ is even (odd). Also note, that there are no zero
modes in this sector for $k<n+1$ or equivalently $q<0$. In particular,
for odd $n$ and vanishing gauge potential there are no zero modes,
in agreement with \cite{Semmelmann:1993}.

For $q\ge0$, the total number of zero modes in the $N\es n$ sector is
\begin{eqnarray}
\label{cpn21}
\sum_{m=0}^{q} \binom{m+n-1}{n-1} = \frac{1}{n!}(q+1)(q+2) \ldots (q+n).
\end{eqnarray}
Similar considerations show that there are no normalizable zero modes
in the $N=0$ sector for $q^\prime<0$, where $q^\prime=\frac{1}{2}(-k-n-1)$.
For $q^\prime \ge 0$ there are zero modes in the $N=0$ sector, and their
number is given by \refs{cpn21} with $q$ replaced by $q^\prime$.

Observe, that the states in the $N=0$ sector are of the same (opposite)
chirality as the states in the $N=n$ sector for even (odd) $n$. The
contribution of the zero modes in those sectors to the index of $\ui\di$
is given by
\eqnl{\frac{1}{n!}(q+1)(q+2)\ldots(q+n), \qquad q = \frac{1}{2}(k-n-1),}{cpn21a} 
for all $q\in\mathbbm{Z}$.

On the other hand, the index theorem on $\C P^n$ reads \cite{Dolan:2002ck}
\eqnl{\ind \ui \di = \int_{\C P^n} \textrm{ch} (L^{-k/2}) \hat A(\C P^n) = \frac{1}{n!}(q+1)(q+2) \ldots (q+n),}{cpn22}
where ch and $\hat A$ are the Chern character and
the $\hat A$-genus, respectively. Note, that this index coincides
with \refs{cpn21a}. This leads us to conjecture, that
for positive (negative) $k$
\emph{all normalizable zero modes} of the Dirac operator on the
complex projective spaces $\C P^n$ with Abelian gauge potential
\refs{cpn8} reside in the sector with $N=n$ ($N=0$) and have the form
\refs{cpn17a}.

We can prove this conjecture in the particular cases $n=1$ and $n=2$.
For $\C P^1$ we have constructed all zero modes. The same holds true
for $\C P^2$ for the following reason: Let us assume that there are
zero modes in the $N=1$ sector.
According to \refs{neo11} they have opposite chirality as compared to
the states in the $N=0$ and $N=2$ sectors. Hence, the index would
be less than the number of zero modes in the extreme sectors.
On the other hand, according to the index theorem, the index
\refs{cpn22} is equal to this number. We conclude that there
can be no zero modes in the $N=1$ sector.

\section*{Conclusions}

We have analyzed $D$-dimensional quantum mechanical systems
that exhibit certain amounts of supersymmetry. Taking the
Hamiltonian to be the square of the Dirac operator, $H = - \di^2$,
on a curved manifold and with background gauge fields,
we have constructed a set of inequivalent `square roots' of $H$.
This set includes, of course, the original Dirac operator as well
as additional supercharges $Q(I_i)$. We have shown how these can be
obtained from complex structures $I_i$. Therefore, the
existence of a higher amount of supersymmetry puts restrictions
on the admissible geometries and gauge configurations. In
even dimensions, $\mN\es 1$ gives no restrictions on the
background fields, while $\mN\es 2$ requires the manifold
to be K\"ahler and the field strength to commute with
the complex structure. The $\mN\es 4$ extended supersymmetry
further requires space to be hyper-K\"ahler and the gauge
field strength to commute with all three complex structures.
In four dimensions this is equivalent to the field strength
being (anti-)selfdual. In $8,12,16,\dots$ dimensions this
requirement is much less restrictive.
In $8$ space dimensions, $\mN\es 8$ has only trivial solutions,
namely flat space without gauge fields. Again, in $16,24,32,\dots$
dimensions there are non-trivial backgrounds admitting
an extended $\mN\es 8$ supersymmetry.

Our construction is similar to the one given in \cite{Rittenberg:1983}. 
However, our approach has the advantage that all objects can be
given a geometric interpretation, like connections, curvature etc. In
addition, for backgrounds admitting extended supersymmetries
(in particular $\mN\es 2$) we can define particle-number
operators $N$ commuting with the super-Hamiltonian (even in curved
space and in the presence of gauge fields).
The complex nilpotent supercharges $Q$ and $Q^\dagger$ act
as lowering and raising operators for the number
operator. The condition $Q^2\es 0$
translates into the existence of a superpotential $g$ for
the (spin)connection as well as for the gauge potential.

As an application, we have deformed the Dirac operator on
$\C P^n$ with the help of $g$ into its free
counterpart and solved the Dirac equation,
for all zero modes of $\ui \di$.

As already mentioned in the introduction,
particular higher dimensional quantum mechanical systems
can be interpreted as supersymmetric field theoretical
models on a spatial lattice. The results obtained
in this paper will turn out to be very useful to
construct supersymmetric sigma-models on a
spatial lattice. This is work in progress, and we
are confident to report on these developments in the near future.

\section*{Acknowledgements}

We thank A. Kotov for many useful discussions, T. Heinzl for
a careful reading of the manuscript and the anonymous referee,
for many helpful comments and for bringing the most interesting
paper \cite{Rittenberg:1983} to our attention. A.K. acknowledges support
by the Studienstiftung des Deutschen Volkes.

\providecommand{\href}[2]{#2}
\begingroup
\raggedright


\endgroup

\end{document}